\documentstyle[prl,twocolumn,aps,psfig]{revtex}
\def\apj #1 #2 #3 {#1, ApJ, {\bf #2}, #3}
\def\apjl #1 #2 #3 {#1, ApJ, {\bf #2}, L#3}
\def\apjs #1 #2 #3 {#1, ApJS, {\bf #2}, #3}
\def\aap  #1 #2 #3 {#1, A\&A, {\bf #2}, #3}
\def\mnras #1 #2 #3 {#1, MNRAS, {\bf #2}, #3}
\def\pra #1 #2 #3 {#1, Phys.~Rev.~A., {\bf #2}, #3}
\def\prb #1 #2 #3 {#1, Phys.~Rev.~B., {\bf #2}, #3}
\def\prc #1 #2 #3 {#1, Phys.~Rev.~C., {\bf #2}, #3}
\def\prd #1 #2 #3 {#1, Phys.~Rev.~D., {\bf #2}, #3}
\def\pre #1 #2 #3 {#1, Phys.~Rev.~E., {\bf #2}, #3}
\def\prl #1 #2 #3 {#1, Phys.~Rev.~Lett., {\bf #2}, #3}
\def\plb #1 #2 #3 {#1, Phys.~Lett.~B., {\bf #2}, #3}
\def\science #1 #2 #3 {#1, Science., {\bf #2}, #3}
\def\nature #1 #2 #3 {#1, Nature., {\bf #2}, #3}
\def\nphysa #1 #2 #3 {#1, Nucl.~Phys.~A., {\bf #2}, #3}
\def\nphysb #1 #2 #3 {#1, Nucl.~Phys.~B., {\bf #2}, #3}
\def\nphysbs #1 #2 #3 {#1, Nucl.~Phys.~B.~Suppl., {\bf #2}, #3}

\def\h#1{\hbox{${}^{#1}$H}}

\def\h502{\hbox{$ h^{2}_{50}$}}

\def\fun#1#2{\lower3.6pt\vbox{\baselineskip0pt\lineskip.9pt
  \ialign{$\mathsurround=0pt#1\hfil##\hfil$\crcr#2\crcr\sim\crcr}}}
\begin{document}
\draft
\title{Constraints on Resonant Particle Production during 
Inflation from the Matter and CMB Power Spectra
}
\author{ 
G.~J.~Mathews$^{1,2}$, D.~J.~H.~Chung$^3$, 
K.~Ichiki$^{2,4}$, T.~Kajino$^{2,4}$, M.~Orito$^{5}$
}
\address{$^1$Center for Astrophysics,
Department of Physics, University of Notre Dame, Notre Dame, IN 46556 }
\address{
$^2$National Astronomical Observatory, 2-21-1, Osawa, Mitaka, Tokyo
181-8588,
Japan}
\address{$^3$Department of Physics, University of Wisconsin, Madison, WI 53706
}
\address{
$^4$University of Tokyo, Department of Astronomy, 7-3-1
Hongo, Bunkyo-ku, Tokyo 113-0033, Japan }
\address{
$^5$Research Laboratory for Nuclear Reactors, 
Tokyo Institute of Technology, Meguro-ku,
Tokyo 152-8550, Japan }
\maketitle
\date{\today}
\begin{abstract}
 We analyze the limits on resonant particle production during
inflation based upon the power spectrum of fluctuations in matter and
the cosmic microwave background.  We show that such a model is
consistent with features observed in the matter power spectrum deduced
from galaxy surveys and damped Lyman-$\alpha$ systems at high
redshift.  It also provides an alternative explanation for the excess
power observed in the power spectrum of the cosmic microwave
background fluctuations in the range of $1000 < l < 3500$.  For our
best-fit models, epochs of resonant particle creation reenter the
horizon at wave numbers of $k_* \sim 0.4 $ and/or $0.2$ ($h$
Mpc$^{-1}$).  The amplitude and location of these features correspond
to the creation of fermion species of mass $\sim 1-2$ M$_{pl}$ during
inflation with a coupling constant between the inflaton field and the
created fermion species of near unity.  Although the evidence is
marginal, if this interpretation is correct, this could be one of the
first observational hints of new physics at the Planck scale.
 \end{abstract}

%
%
%

\section{INTRODUCTION}

Analysis of the power spectrum of fluctuations in the 
the large-scale distribution of matter
(cf.~\cite{cmbreview2,tegmark1,tegmark2}),  
together with fluctuations in the cosmic
microwave background (CMB) 
(cf.~\cite{cmbreview})
 provides powerful constraints on the physics of
the very early universe. 
The most popular account for the
origin of both power spectra is based upon quantum fluctuations 
generated
during the inflationary epoch \cite{cmbinflate}.  
Subsequently, acoustic oscillations of the photon-baryon fluid distort
this to produce the observed features in the angular power spectrum of
temperature fluctuations in the CMB and the spatial power spectrum of
matter density fluctuations.
The two power spectra are different in that the
cosmic microwave background is sensitive to the baryonic material,
while the matter power spectrum probes the dominating dark matter.

\begin{figure}
\psfig{figure=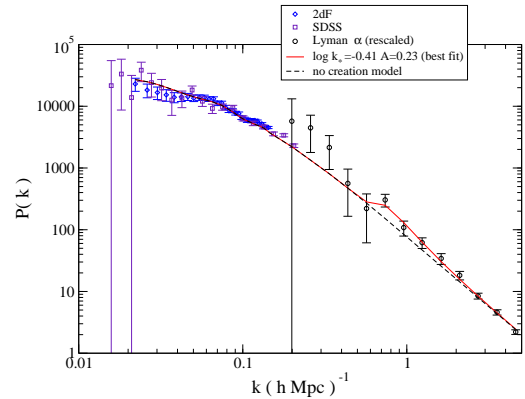,angle=270,width=3.0in}
\vskip .1 in
\caption{Comparison of the observed galaxy cluster function
from  {\it SDSS} [17], {\it 2dF} [18], and Lyman-$\alpha$ [19]
 with the spectrum implied from
the fits to the matter power spectrum with (solid line)  and without (dashed line)
resonant particle creation during inflation as described in the text.}
\label{figpk}
\end{figure}

In this context, 
there now  exist determinations 
of the matter power spectrum on small
angular scales due to  recent 
{\it  SDSS} \cite{tegmark1,sdss}, and {\it 2d}F \cite{2df}
galaxy surveys as well as 
analysis of the Lyman-$\alpha$ forest
\cite{Croft,Gnedin} at higher redshift and even smaller scales.  This
latter  determination is 
particularly facilitated
by the fact that at high
redshift, the Lyman-$\alpha$ absorption systems are still within the
quasi-linear regime so that an 
inference of the primordial power spectrum
is  relatively straightforward. 
The overall matter power spectrum deduced in this way is consistent with
a standard $\Lambda$CDM cosmology, but as shown in Figure \ref{figpk}
 there is marginal evidence of  a peculiar feature beginning near 
$k \sim 0.6$ $h$ Mpc$^{-1}$
 which is not easily explained  away 
by systematic errors \cite{Gnedin}.  There is also 
at least a possibility for
structure to exist in the region near $k \sim 0.3$ $h$ Mpc$^{-1}$.
 In this paper we consider
that such features may be a part of the
primordial spectrum generated by new physics near the end of the
inflation epoch.

Regarding the CMB, data \cite{wmap}
from {\it WMAP} have placed stringent constraints on the
cosmological parameters relevant to the observed power 
spectrum in the range
of multipoles up to $l \le 1000$.  
The simplest prediction of the standard $\Lambda CDM$ cosmology
is that the primordial power spectrum should be strongly damped at 
higher multipoles
due to photon diffusion (Silk damping) and because more than one
perturbation can fit within the depth of the surface of last
scattering, and hence, be washed out.

\begin{figure}
\psfig{figure=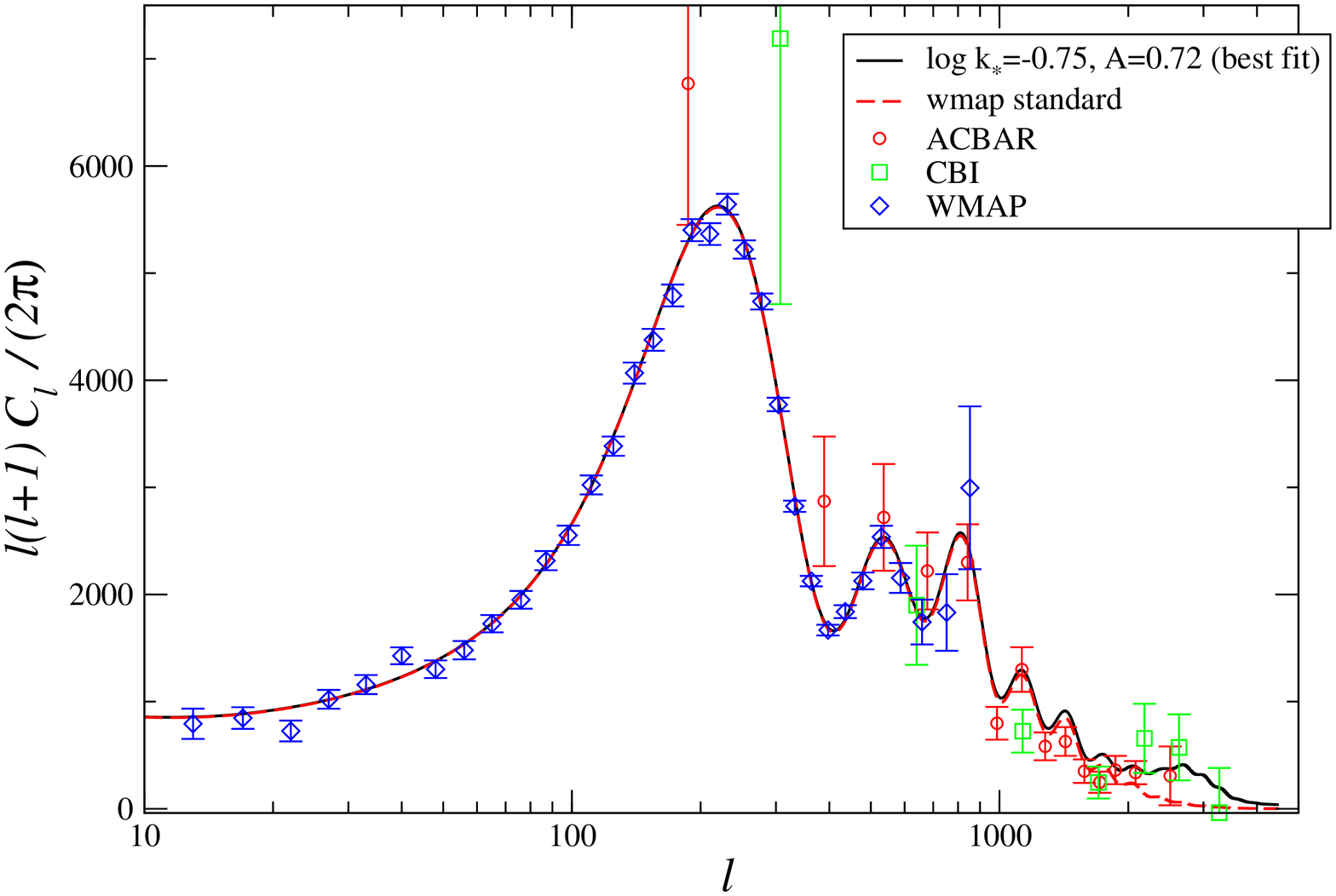,angle=270,width=3.0in}
\psfig{figure=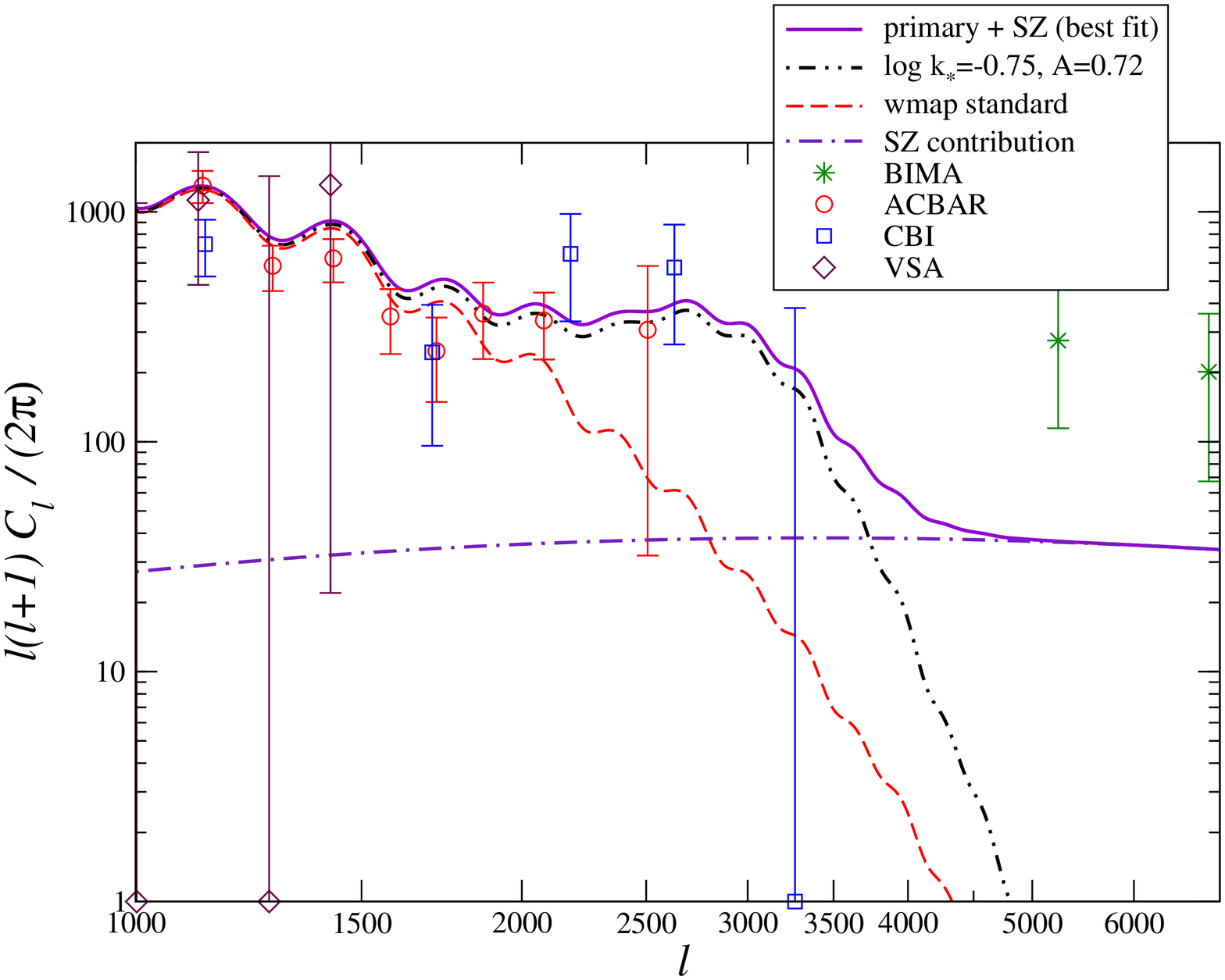,angle=270,width=3.0in}
\vskip 0.3 in
\caption{ Upper figure shows the CMB {\it WMAP, ACBAR,} and {\it CBI} 
data in the range $l = 10 - 5000$.  The dashed line
is the CMB power spectrum computed using 
the standard {\it WMAP} cosmological parameters.  
The thick solid line is for a best-fit to these data for a model  
with resonant particle production included.
The lower figure shows 
an expanded view of the {\it ACBAR, CBI, VSA}, and {\it BIMA}
data in the range of $l = 1000 - 7000$.
The dashed line shows the standard {\it WMAP} result without the SZ effect included.
The dot-dashed line shows our fit  SZ contribution
(from the analytic halo model of Ref.~[14]).
The solid line shows the best fit to these data sets  with resonant particle
production and the SZ effect included. The dot-dot-dashed line shows
the resonant particle creation component  without the SZ contribution.
}
\label{cmbfig}
\end{figure}

Of interest for the present work, however,  is the recent accumulation
of  observations of the
CMB power spectrum in the range $1000 < l < 7000$ 
by various groups
({\it CBI} \cite{cbi1,cbi2,cbi3}, {\it ACBAR} \cite{acbar},
{\it BIMA} \cite{BIMA}, and {\it VSA} \cite{VSA}).
These observations on the smallest angular scales 
 are of particular interest as a test of this basic prediction
of the standard inflation/photon-decoupling
paradigm. The current data are summarized in Figure 
\ref{cmbfig} from which it is clear that the deduced 
power spectrum increases
rather than decreases for large multipoles, particularly in the range 
$2000 < l < 3500$.

The most likely interpretation (cf. \cite{Bond,Komatsu})
of an excess power at high multipoles is
a manifestation of the scattering of CMB photons 
by hot electrons in clusters known as 
the thermal Sunyaev-Zeldovich (SZ) effect \cite{SZ}.
Although some contribution from the SZ effect undoubtedly occurs,
it is not yet conclusively  established 
that this is the only possible interpretation 
(cf. \cite{Aghanim}) of excess power on small angular scales.
Indeed, it has been deduced \cite{Bond,Komatsu} that 
explaining the excess power in the observed CMB spectrum
requires that  the mass density fluctuation
amplitude parameter,  $\sigma_8$, be near the upper end
of the range of the values deduced via other independent means.
Another significant effect is the large cosmic variance.
The uncertainties on the SZ contribution are highly non-Gaussian. 
They thus depend strongly upon whether or not the observed 
field happens to contain a group
or cluster of galaxies.  Hence, there is large uncertainty in any deduced
SZ effect.
In view of these uncertainties, an exploration of other possible explanations for
this power excess seems warranted.

Indeed, alternative explanations exist
for the generation of features in both  the matter and CMB power
spectra on small angular scales.  
For example, a flattening of the
inflation generating effective potential near the end of inflation
could produce such distortions on small angular scales \cite{hannu}.

In this paper, however, we consider another alternative originally
proposed in \cite{chung00} whereby such distortions might arise from
the resonant production of particles during large field
inflation.\footnote{Large field inflation is a slow roll inflationary
scenario in which the inflaton scalar field has a classical field
value exceeding the Planck scale.}  This interpretation has the
intriguing aspect that, if correct, an opportunity emerges to use the
CMB and matter power spectra as probes of the Planck scale ($M_{pl} \sim
10^{19}$ GeV) particle spectrum.

The prototypical scenario is that there is an inflaton $\phi$ which
controls the mass of a fermion $\psi$ or a boson $\chi$ through the coupling 
\begin{equation}
L_{int}= -[M_f - M_{pl} f(\frac{\phi}{M_{pl}})] \bar{\psi} \psi - [M_B^2 -
  M_{pl}^2 g(\frac{\phi}{M_{pl}})] \chi^2 ,
\end{equation}
where $M_f$ and $M_B$ are fermion and boson masses, respectively,
which are assumed to be of order $M_{pl}$.  Since by the definition of
large field inflationary scenario $\phi/M_{pl} \gtrsim 1$ and
$\phi/M_{pl}$ varies at least an order of magnitude, any natural order
unity functions $f(x)$ and $g(x)$ will lead to a cancellation of the masses
during the evolution of slow roll inflation.  Due to this
cancellation, when $|M_f - M_{pl} f| \lesssim H$ and/or $|M_B - M_{pl}
g| \lesssim H$, the effective mass of $\psi$ and $\chi$ is varying
nonadiabatically (since the inflaton field variation time rate is also
typically not too far from the Hubble expansion rate
$H$).\footnote{More precisely, for quadratic monomial inflaton
potentials, the nonadiabaticity condition $\frac{\dot{w}_k}{w_k^2} >1$
leads approximately to \[H > y (M_f- M_{pl}g)\] for fermion mass term
where \[y= \frac{4\pi |g^{-1}(M_f/M_{pl})|}{|g'|} \left( 
\frac{M_f}{M_{pl}} - g\right). \]} During this nonadiabatic
period, there is 
efficient particle production which can be thought of as a kind of
resonant particle production.

In \cite{chung00}, the effect of the resonant {\em fermionic} particle
production was taken into account neglecting the nonadiabatic effects
on the modes outside of the horizon.  This leads to a bump-like
structure in the primordial power spectrum.  However,
\cite{Elgaroy:2003hp} considered the nonadiabatic effects of bosonic
particle production on the modes outside the horizon.  Their claim is
that the primordial power spectrum is modified by a steplike structure
rather than a bumplike structure.  Given that their analysis was for
bosonic particle production while we are going to deal with fermionic
production following \cite{chung00},\footnote{Dealing with fermionic particle production allows
us to avoid dealing with nonminimal phase transition effects that can
arise due to the possible nontrivial ground states of the bosonic
field.} and given that the subtle effect of \cite{Elgaroy:2003hp}
(which was only approximately computed numerically) requires more
detailed attention to unambiguously establish the change in the
qualitative behavior of the primordial spectral feature, we will in
this paper consider only the semianalytic results of \cite{chung00}
when fitting to data.  However, because of this omission of the
nonadiabatic effect, this result should only be considered indicative
of the type of constraints one can obtain from the latest data.  A
detailed analysis of the nonadiabatic effects will be considered in a
future publication.

Planck-scale mass particles generically exist in Planck
scale compactification schemes of string theory from the Kaluza-Klein states,
winding modes, and the massive (excited) string modes.
Hence, the existence of Planck-scale
mass particles which couple to the inflaton is a generic situation.  What
is perhaps not generic in the scenario considered in this paper is
that the mass of the Planck-scale particle lies in the $e$-fold range
of the inflaton accessible to observation. 

Even with all of these favorable assumptions, we find only marginal
hints for the existence of such a particle from fits to current data.
More specifically, we find marginal hints for a bump at a scale of
$k\sim 0.4$ (and 0.2) $ h$ Mpc$^{-1}$ in the primordial 
spectrum due to production
of fermions of mass $m \approx 2 M_{pl}$ with an $O(1)$ weak coupling
to the inflaton.

The order of presentation will be as follows.  In section II, we
briefly review and clarify the semianalytic results of \cite{chung00}
and then set up the parameterization of the primordial spectrum
fitting function.  In sections III and IV, we briefly discuss the
matter power spectrum parameterization and the CMB power spectrum
(including the SZ effect which is the more conservative interpretation
of the excess power at large $l$ values). Section V is devoted to the
discussion of the fitting procedure and the fit results to the matter
and the CMB spectrum.  Finally, we conclude in section VI.

\section{Inflation Resonant Particle Production}

In the basic inflationary picture, a rapid early expansion of the
universe is achieved through the vacuum energy from an inflaton field.
In the minimal extension from the basic picture considered here, the
inflaton is postulated to couple to at least one massive particle
whose mass is order of the of the inflaton field value.  This particle
is then resonantly produced as the field obtains a critical value
during inflation.  If even a small fraction of the energy in the
inflaton field is extracted in this way, it can produce features in
the primordial power spectrum. In particular, there will be excess
power in the spectrum at the angular scale corresponding to when the
epoch of resonant particle creation crossed the Hubble radius.

In the simplest slow roll approximation \cite{cmbinflate} for the
generation of density perturbations during inflation, the amplitude,
$\delta_H(k)$, of a density fluctuation when it crosses the Hubble
radius is just,
\begin{equation}
\delta_H(k) \approx {H^2
\over 5 \pi \dot \phi}~~,
\label{pert}
\end{equation}
where $H$ is the expansion rate, and $\dot \phi$ is the velocity of
the inflaton field when the comoving wave number $k$ crosses the
Hubble radius during inflation.  If resonant particle production
drains energy from the inflaton field, then the conjugate momentum in
the field $\dot \phi$ decreases.  This causes an increase in
$\delta_H(k)$ (primordial power spectrum) for those wave numbers which
exit the horizon during the resonant particle production epoch.

Of course when $\dot{\phi}$ is changing due to particle production,
$\ddot{\phi}$ may not be negligible, resulting in corrections to
Eq.~(\ref{pert}).  In \cite{chung00}, this correction was considered
and found to be $<<20 \%$ \footnote{This fraction
  refers to the fraction of the particle production effect, not the
  entire power spectrum amplitude.}  for the particle
production of interest for the fits of this paper.

The inflaton field is then postulated to have a simple Yukawa coupling
to a fermion field $\psi$ of mass $m$ in the form,
\begin{equation}
{\cal L}_Y = -\lambda \phi \bar \psi \psi~~.
\label{eq:yukawacoupling}
\end{equation}
Including this new coupling, the equation of motion
for the inflaton field becomes
\begin{equation}
\ddot \phi + 3H \dot \phi + {dV \over d\phi} - 
N \lambda  \langle \bar \psi \psi \rangle = 0 ~~,
\label{eom}
\end{equation}
for $N$ fermions of mass $m$ coupled to the inflaton.
The effective mass of the fermion is
$M(\phi) = m - \lambda \phi$, which vanishes
for a critical value of the inflaton field,
$\phi_* = m/\lambda$.  Resonant fermion production
will then occur in a narrow range of inflaton field amplitude
around $\phi = \phi_*$.

As in \cite{chung00} we label the epoch at which particles are created
by an asterisk.  So the cosmic scale factor is labeled $a_*$ at the
time $t_*$ at which resonant particle production occurs.  Considering
a small interval around this epoch, one can treat $H = H_*$ as
approximately constant (slow roll inflation).  The number density $n$
of particles can be taken as zero before $t_*$ and afterwards as $n =
n_*[a_*/a(t)]^{3}$.  The fermion vacuum expectation value can thus be
written,
\begin{eqnarray}
\langle \bar \psi \psi \rangle & \approx & n_*~\theta(t-t_*)[a_*/a(t)]^{3} 
\nonumber \\
&&
\approx n_*~\theta(t-t_*)\exp{[-3H_*(t-t_*)]}~~.
\end{eqnarray}

Now inserting this relation into the equation of motion
for the inflaton field (Eq.~(\ref{eom})), one can obtain
the change in the inflaton field evolution
 $\dot \phi$ due to particle
creation,
\begin{eqnarray}
\dot \phi(t > t_*) &=& \phi(t > t_*)_{\lambda = 0} 
\nonumber \\
&+& N \lambda n_* \theta(t-t_*)\exp{[-3H_*(t-t_*)]}~~.
\label{dotphi}
\end{eqnarray}

Inserting this into Eq. (\ref{pert}), 
a very good analytic approximation to the
effect of the particle creation on the
perturbation spectrum can be obtained \cite{chung00},
\begin{equation}
\label{eq:deltaanalytic}
\delta_H(k) =  \frac{ [\delta_H(k)]_{\lambda=0}}
{1-\theta(a-a_*) |\dot{\phi}_*|^{-1}
N\lambda n_* H_*^{-1}(a_*/a)^3\ln(a/a_*)}~~ .
\end{equation}
The scale factor $a$ relates \cite{Liddle}
 to the physical wave number $k$ by,
\begin{eqnarray}
\label{Ncross}
\ln \frac{k}{a_0 H_0} &=& 62 + \ln \biggl[\frac{a}{a_*}\biggr]
+  \ln \biggl[\frac{a_*}{a_{\rm end}}\biggr]
\nonumber \\
&&
        - \ln \frac{10^{16}\,{\rm GeV}}{V_k^{1/4}}
        + \ln \frac{V_k^{1/4}}{V_{{\rm end}}^{1/4}} - \frac{1}{3} \ln
        \frac{V_{{\rm end}}^{1/4}}{\rho_{{\rm reh}}^{1/4}}~~ ,
\end{eqnarray}
where $a_0H_0 \approx (h/3000)$ Mpc$^{-1}$ denotes the present 
comoving Hubble scale.  The subscript `$k$'
indicates the inflaton effective potential
value when a particular wave number $k$ crosses the Hubble radius
during inflation ($k=aH$).  The quantities $a_{end}$ and $V_{end}$
 are the scale factor and effective inflaton potential at
the end of inflation, and $\rho_{{\rm reh}}$ is the 
matter energy density after reheating 
to the standard hot big bang Friedmann cosmology.
This expression assumes that instantaneous
transitions occur between the various regimes, and that the
universe behaves as if matter-dominated during reheating.

Using this relation between scale factor and $k$,
the perturbation spectrum (Eq.~(\ref{eq:deltaanalytic})) 
can be reduced \cite{chung00} to a simple 
two-parameter function.
\begin{equation}
\label{eq:fit}
\delta_H(k) = \frac{\left[\delta_H(k)\right]_{\lambda=0}}
{1-\theta(k-k_*) A(k_*/k)^3\ln(k/k_*)}~~~,
\end{equation}
where the coefficient $A$ and characteristic wave number $k_*$ ($k/k_*
\ge 1$) can be fit to the observed power spectra.  (Note that this $A$
is {\em different} from the $A$ coefficient of \cite{chung00}.)

The values of $A$ and $k_*$ determined from observation directly
relate to the inflaton coupling $\lambda$ and fermion mass $m$, for a
given inflation model.  When the back reaction is not important, we
can write
\begin{equation}
A  = |\dot{\phi}_*|^{-1} N\lambda
n_* H_*^{-1} 
\end{equation}
which uses the approximation
\cite{chung00,birrellanddavies,Kofman:1997yn,Chung:1998bt} for the
particle production Bogoliubov coefficient to be
\begin{equation}
|\beta_k|^2 = \exp\left( \frac{-\pi k^2}{a_*^2 \lambda |\dot \phi_*| }\right).
\end{equation}
Note that this approximation does not depend on the particular form of
the inflaton potential.  Its main assumptions are only that the
particle mass being produced is nearly negligible during the resonant
production (which is an excellent assumption for our case) and that
the effective-mass time variation due to the inflaton time variation
dominates over the contribution due to the FRW expansion.  Hence, when
we carry out the fits, we can approximately marginalize over the
spectral index (since each constant spectral index corresponds to a
different inflationary model in which the spectral index is
approximately constant).

The coefficient $A$ in this approximation of negligible back reaction
can be related directly to the coupling constant $\lambda$ by noting
that
\begin{equation}
\label{eq:nstar}
n_* = \frac{2}{\pi^2}\int_0^\infty dk_p \, k_p^2 \, 
|\beta_k|^2 =
\frac{\lambda^{3/2}}{2\pi^3}
|\dot{\phi}_*|^{3/2}~~ .
\end{equation}
This give us
\begin{eqnarray}
A & = & \frac{N \lambda^{5/2}}{2 \pi^3}
\frac{\sqrt{|\dot{\phi}_*|}}{H_*}\\
& \approx & \frac{N \lambda^{5/2}}{2 \sqrt{5} \pi^{7/2}}
\frac{1}{\sqrt{\delta_H(k_*)|_{\lambda=0}}}
\label{eq:alamrelation}
\end{eqnarray}
where we have used the usual approximation for the primordial slow
roll inflationary spectrum \cite{cmbinflate}.  This means that
regardless of the exact nature of the inflationary scenario, for any
fixed inflationary spectrum $\delta_H(k)|_{\lambda=0}$ without the
back reaction, we have the particle production giving us a bump of the
form Eq.~(\ref{eq:fit}) with the parameter $A$ expressed in terms of
the coupling constant through Eq.~(\ref{eq:alamrelation}).  Given that
the CMB normalization requires $\delta_H(k)|_{\lambda=0}\sim
10^{-5}$, we have 
\begin{equation}
A \sim 1.3 N \lambda^{5/2}.
\end{equation}
Hence, for $A \sim O(0.1)$, both $\lambda<1$ and $\lambda N<1$ are
possible, satisfying perturbativity.  As will be seen in the sections
below, our best fits will indeed give $A \sim O(0.1)$.  When $N$ and
$\lambda$ are sufficiently large that the back reaction becomes
important (when the naively computed $A \sim O(3)$),\footnote{The back
reaction becomes very strong as $A \rightarrow 3 e \approx 8$.} the
back reaction reduces \cite{chung00} the actual amplitude of the peak
relative to the perturbative value given by
Eq.~(\ref{eq:alamrelation}).

Before we conclude this section, we would like to explicitly state
further caveats to using Eq.~(\ref{pert}) in computing the primordial
perturbation spectrum:
\begin{enumerate}

\item Instead of the usual long wavelength gauge invariant
curvature perturbation variable implicit in using Eq.~(\ref{pert}),
the gauge invariant gravitational potential must be recomputed
incorporating the coupling of  Eq.~(\ref{eq:yukawacoupling}).

\item Even with the usual gauge invariant curvature perturbation
  formalism, Eq.~(\ref{pert}) neglects the change in pressure that
  occurs due to the particle production.  Even considering the
  adiabatic perturbation component, this would change the denominator
  of Eq.~(\ref{pert}).  

\item The numerator of Eq.~(\ref{pert}) also obtains a  contribution from
  the particle production which we are neglecting.

\item As pointed out by \cite{Elgaroy:2003hp}, perhaps the most
  significant effect that has been neglected is the nonadiabatic
  pressure change.

\end{enumerate}
A more detailed investigation of these neglected effects will be
deferred to a future publication.

\section{Matter Power Spectrum }

It is straight forward to determine the matter power spectrum to
compare with that deduced from large-scale structure surveys
\cite{sdss,2df} and the Lyman-$\alpha$ forest \cite{Croft}.  To
convert the amplitude of the perturbation as wave number $k$ enters
the horizon, $\delta_H(k)$, to the present-day power spectrum, $P(k)$,
which describes the amplitude of the fluctuation at a fixed time, one
must make use of a transfer function, $T(k)$ \cite{efstathiou} which
is easily computed using the code CMBFAST \cite{cmbfast} for various
sets of cosmological parameters (e.g.~$\Omega$, $H_0$, $\Lambda$,
$\Omega_B$).  An adequate approximate expression for the structure power
spectrum is
\begin{equation}
\frac{k^3}{2\pi^2}P(k) = \left( \frac{k}{aH_0} \right)^4 T^2(k)
\delta^2_H(k) \ .
\end{equation}
This expression is only valid in the linear regime,
which in comoving wave number is up to approximately $k ^<_\sim
0.2~h$ Mpc$^{-1}$ and therefore
adequate for our purposes.
However, we also correct for
the nonlinear evolution of the power spectrum \cite{Peacock}.

\section{CMB power spectrum}
Features in the primordial power spectrum will also appear in the
observed  CMB
temperature fluctuations.  The connection between the
resonant particle creation and CMB temperature fluctuations 
is straightforward.  As usual, temperature fluctuations 
are expanded in spherical harmonics, $ \delta T/T =\sum_l\sum_m
a_{lm}Y_{lm}(\theta,\phi)$ ($2 \le l<\infty$ and $-l \le m \le l$). 
The anisotropies are then described by the angular 
power spectrum, $C_l= \langle |a_{lm}|^2\rangle$, as 
a function of multipole number $l$.  One then merely
requires the conversion from perturbation spectrum $P(k)$
to angular power spectrum $C_l$.  This is also easily 
accomplished using the code CMBFAST \cite{cmbfast}.
As input to CMBFAST 
we adopt the usual power law primordial spectrum plus the
perturbation due to resonant particle production.
For speed, CMBFAST does not compute all $C_l$, but uses a spline
fit to interpolate.  We have checked the stability of the CMBFAST results
when adding such features to the power spectrum by increasing the
number of $C_l$ explicitly computed.   The results were convergent
even for the default spline fits of CMBFAST.
When converting to the angular power spectrum,
the amplitude of the narrow particle creation
feature  in $\delta_H(k)$ is spread over many values of $l$.
Hence, the particle creation feature looks like a broad peak which is
easily accommodated even when implementing spline fits.

\subsection*{SZ effect}
It has been proposed \cite{Bond,Komatsu} that 
the observed spectrum at high $l$ is explained by the
SZ effect.  However, 
the amplitude of the SZ contribution to the
power spectrum is very sensitive to the 
parameter $\sigma_8$ describing rms mass fluctuation
within a fiducial 8 $h^{-1}$ Mpc sphere.
The amplitude of the expected SZ peak
scales as $\sigma_8^7$.

Explaining the excess power in the observed CMB spectrum
requires that  the mass density fluctuation
amplitude parameter,  $\sigma_8$, be  slightly above unity
\cite{Komatsu}. 
However, a variety of independent measures 
including weak lensing \cite{Brown}, galaxy velocity fields
\cite{Willick}, galaxy clusters at high redshift \cite{Bahcall},
X-ray emitting clusters \cite{Voevodkin},
and the independent value 
obtained by the {\it WMAP} fit \cite{wmap} to the
lower multipole data all favor a mean value in the range
$\sigma_8 \sim 0.7 - 0.9$ (see Ref.~\cite{Bond} for a recent review).
Note that even a slight  reduction of
the amplitude parameter by 10\% is sufficient to
reduce the magnitude of distortion  by more than a factor of
two.  For the present purposes we adopt a conservative value of
$\sigma_8 = 0.9 \pm 0.1$ as a reasonable
prior distribution based upon the various independent measures
of $\sigma_8$.  
 
 In what follows we fit the 
amplitude of the SZ contribution using 
the SZ power spectrum calculated in
\cite{Komatsu} based upon the analytic halo formalism
\cite{Cole}.  
This analytic form has been shown \cite{Bond,Komatsu} to
adequately represent the power spectra deduced
from numerical simulations.

\section{results}
We have made a  multi-dimensional 
 Markov Chain Monte-Carlo
analysis \cite{Christensen,Lewis} of the 
mass power spectrum based upon the combined $2dF$, $SDSS$, and
Lyman-$\alpha$ data
in models with and without resonant particle creation during
inflation to alter the primordial fluctuation spectrum.
We also have independently analyzed the CMB using the combined
{\it WMAP, CBI, ACBAR, } and {\it VSA} data.  In addition,
we made a total analysis based upon
the combined mass and CMB power spectra.  

For simplicity and speed in the present study we
only marginalized over the five
parameters which do not alter the matter or CMB transfer functions.
Hence, the set of free parameters in the analysis is ($n_s, A_s, \log{(k_*)},
A, A_{SZ}  $), where $n_s$ is the spectral index, 
$A_s$ is the overall amplitude of the primordial power spectrum,
and $A_{SZ}$ is the overall amplitude of the SZ effect which we relate
to an effective $\sigma_8^{SZ}$ parameter for the SZ effect.
The true $\sigma_8$ parameter, however,  is determined by $A_s$ once the
power spectrum and transfer functions are fixed.
As usual,
both $n_s$ and $A_s$ are normalized 
at $k = 0.05$ Mpc$^{-1}$. 
As noted above, we 
adopt a  conservative prior of $\sigma_8^{SZ} = 0.9 \pm 0.1$ 
 as opposed to the best fit combined
{\it WMAP} analysis value of $\sigma_8 = 0.84 \pm 0.04$.  
For this illustrative study all other parameters for this 
analysis were fixed at the
optimum {\it WMAP} parameters \cite{wmap},
i.e. ($h, \Omega_b h^2, \Omega_m h^2, \Omega_\Lambda,  \tau ) = 
(0.71, 0.0224, 0.135, 0.75, 0.17)$.  One expects that different
results would arise by optimizing over other parameters,
e.g. $h, \Omega_b h^2$, etc.
However, for the present exploratory study, we only wish to
identify angular scales which might be of interest for future study.
For this purpose, a simple five parameter marginalization
is adequate since such parameters do not produce the
kind of spectrum bump of interest here.

\subsection{Matter power spectrum fit}

\begin{figure}
\psfig{figure=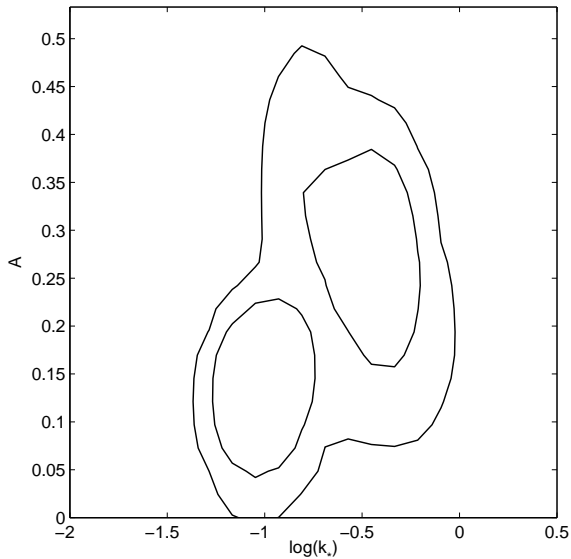,width=3.0in}
\caption{Constrains on parameters $A$ and $k_*$ from the
fit to the matter power spectrum alone.
Contours show 1 and 2 $\sigma$ confidence limits. The horizontal
axis indicates $\log (k_*)$ where $k_*$ is in units of ($h$ Mpc$^{-1}$). }
\label{contourfigpz}
\end{figure}

 Contours of constant goodness of fit in the $A$ vs.~$k_*$ 
plane consistent with the matter power spectrum constraint 
are shown on Figure \ref{contourfigpz}.
These fits are based upon the {\it 2dF}, {\it SDSS}, and Lyman-$\alpha$ data
described above. Window functions required for fits to
these data sets are given in approximate analytic form 
in \cite{2df} for the {\it 2dF} data,
and in \cite{sdss} for the {\it SDSS} data.  
As described in \cite{verde}, a simple $\chi^2$  is used to fit 
for the Lyman-$\alpha$ data, because
the full correlation matrix is not available.
 
  The most recent {\it SDSS} matter power spectrum extends into
the nonlinear regime, but is probably not reliable \cite{tegmark1} for
the last band widths above $k = 0.2~ h$  Mpc$^{-1}$.
Hence, we have omitted  the
last points of the {\it SDSS} 
for $k \ge 0.2~ h$  Mpc$^{-1}$
 and 
for $k \ge 0.15 ~ h$  Mpc$^{-1}$
for the {\it 2dF} power spectrum
 as recommended \cite{tegmark1}.
For the {\it SDSS} and {\it 2dF} power spectra
the unknown bias factors (multiplier to get from the matter power
spectrum to the galaxy power spectrum) 
 were analytically marginalized 
according to the method described in the appendix of Ref.~\cite{Lewis}.

The contours on Figure \ref{contourfigpz} identify two possible regions which
could be consistent with resonant particle production.
The feature with the largest amplitude
occurs at $k_* = 0.41 \pm 0.05~h$\,Mpc$^{-1}$
and with $A = 0.23 \pm 0.08$. These parameters correspond to 
$m \approx  2.2 $ M$_{pl}$ and $\lambda \approx  0.6$ for a 
single fermion species.

The weaker feature has a minimum at $k_* = 0.08 \pm 0.01~h$ Mpc$^{-1}$ and 
an amplitude $A = 0.12 \pm 0.04$.  This feature is more or less  consistent
with a similar feature seen in the CMB, 
but at a lower amplitude and slightly larger  angular scales.
This feature would correspond to $m \approx  1.8$ M$_{pl}$ 
and $\lambda \approx  0.5$.
 However, it occurs at the interface between 
the SDSS and Lyman-$\alpha$ power spectra, and hence, may be an artifact of the 
matching of the data sets rather than a real bump in the spectrum.  Obviously, this is
another  feature which warrants further careful scrutiny.

Figure \ref{figpk} shows the effect of the stronger feature on the
matter power spectrum compared with data from {\it 2dF}, {\it SDSS},
and Lyman-$\alpha$. Clearly the feature in the matter power spectrum
beginning near $k \approx 0.6~h$ Mpc $^{-1}$ is important in this regard and
warrants careful scrutiny.  (Note that the peak of the primordial
spectrum bump generically occurs at $k \approx \exp(1/3) k_*$ with our
parameterization.)

\subsection{CMB Fit}

\begin{figure}
\psfig{figure=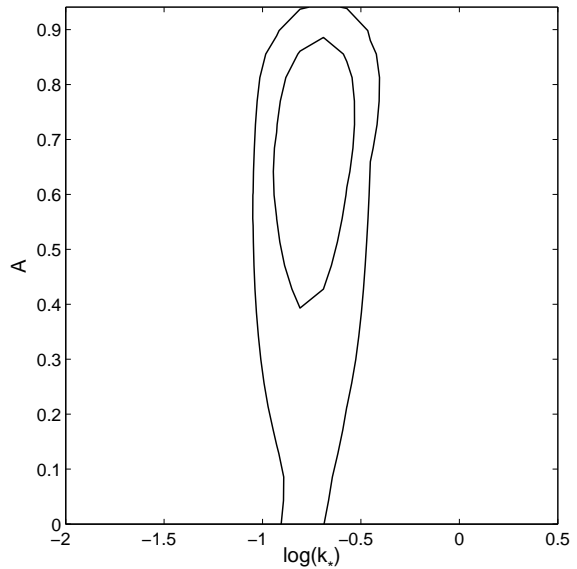,width=3.0in}
\caption{Constrains on parameters $A$ and $k_*$ from the
CMB power spectrum.
Contours show 1 and  2$ \sigma$ limits.  The horizontal
axis indicates $\log (k_*)$ where $k_*$ is in units of ($h$ Mpc$^{-1}$). }
\label{contourfig}
\end{figure}

 Cosmological parameters with and
 without resonant particle creation were obtained 
 from the combined 
{\it WMAP, CBI, ACBAR, } and {\it VSA} 
data using a Markov chain Monte-Carlo
analysis \cite{Christensen} as described above. 
Figure \ref{contourfig} shows the 1 and 2$\sigma$ contours 
in the $k_*$ vs.~$A$ plane from this analysis.
The CMB power spectrum 
 is best fit for 
$A = 0.7 \pm 0.2 $ and $k_* = 0.18 \pm 0.02~ h$ Mpc$^{-1}$.  

 Figure \ref{cmbfig} shows our best fit to the CMB
data  for models with and without
resonant particle creation.
Also shown for comparison in the lower expanded figure is the SZ contribution
for $\sigma_8 = 0.9$.
We find that an excellent fit to the ACBAR and CBI  
observed CMB power spectra can be achieved in this way.  
Indeed, those  CMB observations
favor this interpretation over the 
conventional SZ plus $\Lambda CDM$ models
at the level of $5\sigma$. However, the {\it BIMA} data set  at
$l \approx 5000 - 7000$ is  a  crucial test of this
possible interpretation.  If the power spectrum is as high
at these l-values as the {\it BIMA} data imply, then an SZ interpretation 
with a large $\sigma_8$ is favored.

\subsection{Combined Analysis}
\begin{figure}
\psfig{figure=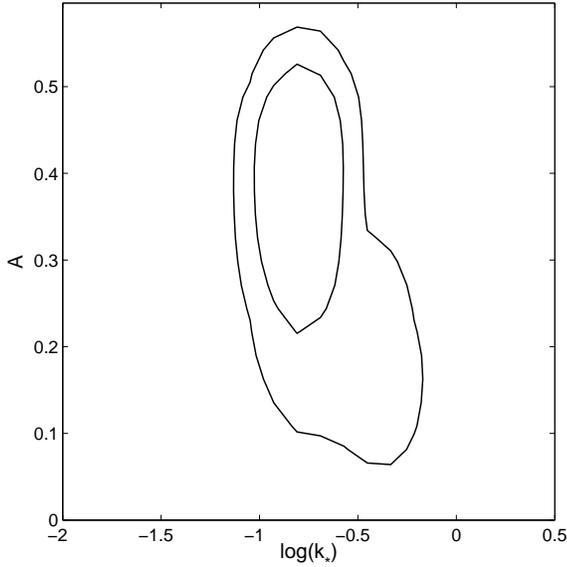,width=3.0in}
\caption{Constrains on parameters $A$ and $k_*$ from 
a fit to the combined 
CMB  and matter  power spectra.
Contours show 1  and $2 \sigma$ limits.  
 The horizontal
axis indicates $\log (k_*)$ where $k_*$ is in units of ($h$ Mpc$^{-1}$).}
\label{contourfigtot}
\end{figure}

\begin{figure}
\psfig{figure=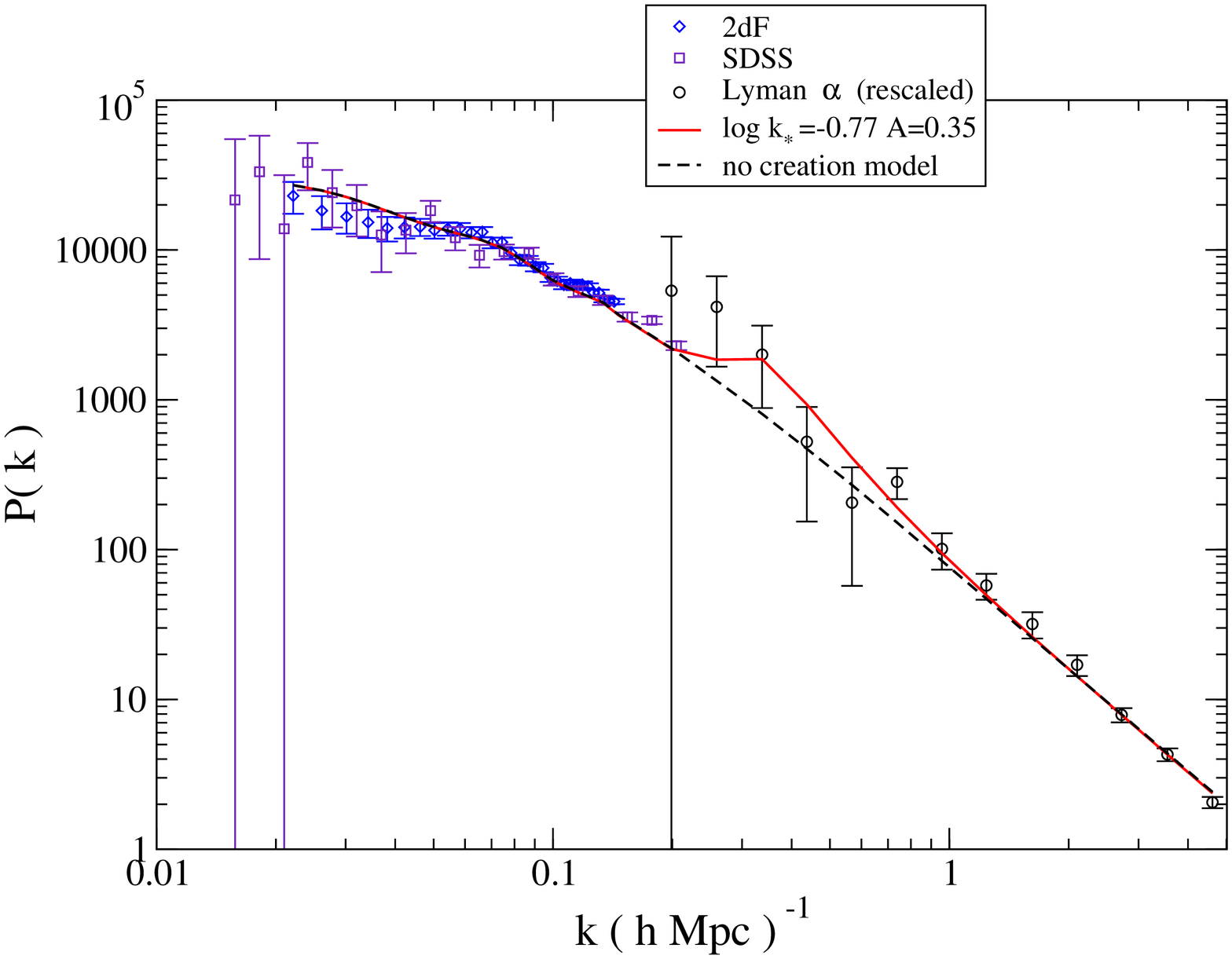,angle=270,width=3.0in}
\psfig{figure=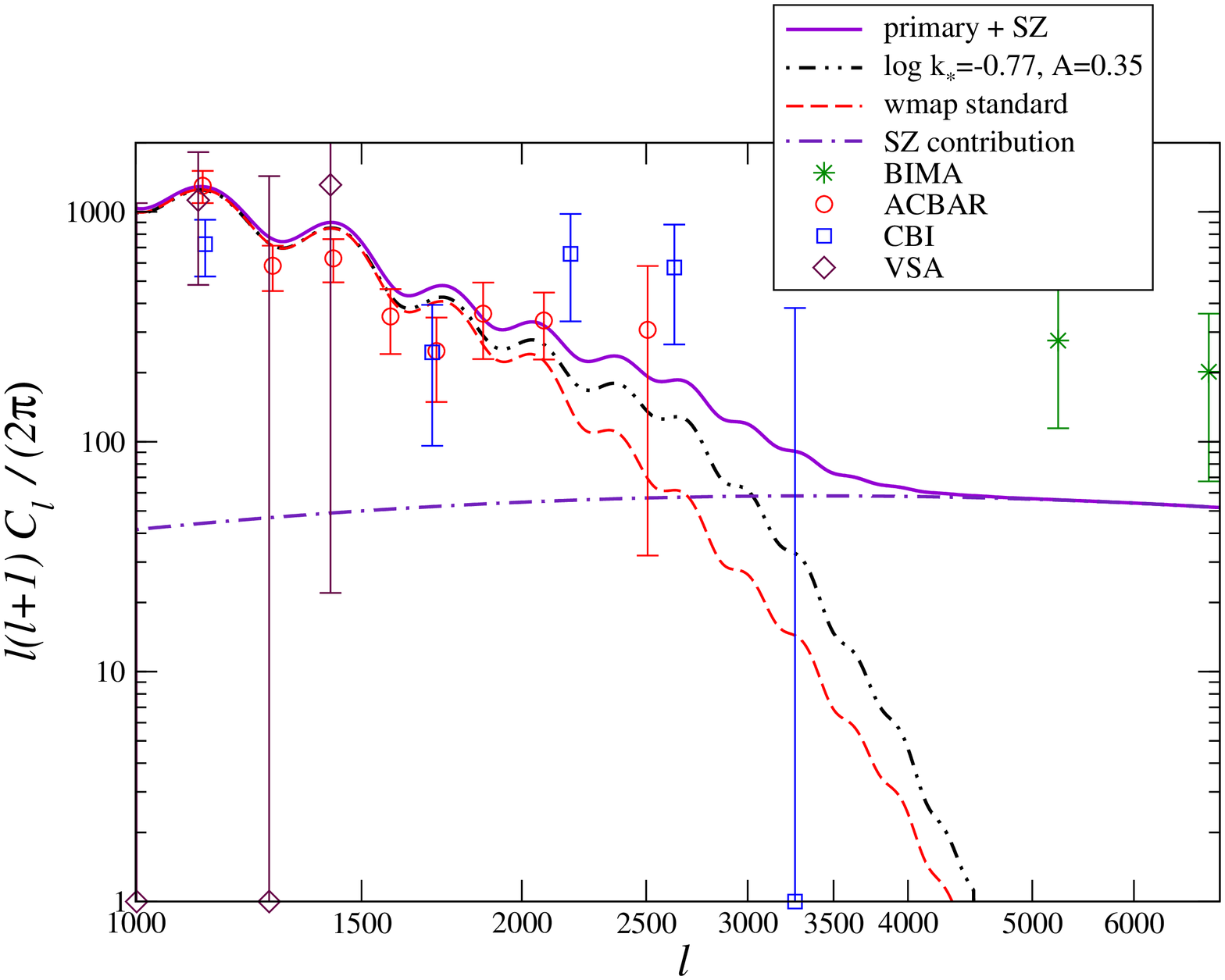,angle=270,width=3.0in}
\vskip .1 in
\caption{ Optimum fit to the combined CMB + matter power spectra.
Upper figure shows the observed galaxy cluster function
from {\it SDSS} [17], {\it 2dF} [18], and Lyman-$\alpha$ [19]
data.  The solid line shows  the spectrum implied from
the optimum model with 
resonant particle creation during inflation.
The lower figure shows
an expanded view of the {\it ACBAR, CBI, VSA}, and {\it BIMA}
data in the range of $l = 1000 - 7000$.
The dashed line shows the standard {\it WMAP} result without the SZ effect included.
The solid line shows the optimum fit with resonant particle
production and the SZ contribution. The d0t-dot-dashed
line shows the resonant particle creation component without the SZ effect.}
\label{fitcomb}
\end{figure}

Contours of $A$ and $k_*$ consistent
with the combined CMB and matter power-spectrum constraints
are shown on Figure \ref{contourfigtot}.
Here it is apparent that a single feature which begins in the
matter power spectrum  at 
$k_* = 0.17 \pm 0.04~h$\,Mpc$^{-1}$ and has an amplitude of
 $A \approx  0.35 \pm 0.10$ is 
most prominent.  Figure \ref{fitcomb} illustrates the associated optimum
fits to the
matter power spectrum (upper curve) and the CMB (lower curve).

As remarked above, even though a feature with $k_* \sim 0.2$
 appears in both
the matter and CMB power spectra, it is not unambiguously attributable
to resonant particle production.  In the
matter power spectrum it could be an artifact
of the uncertainty in joining of  the galaxy and Lyman-$\alpha$
data sets near $k \approx 0.1$,
while  in the CMB its significance could be confused by the SZ effect.
Nevertheless, the feature with $k_* = 0.4 ~h$\,Mpc$^{-1}$ and $A \approx  0.2$
which is clearly seen in the
matter power spectrum of Figure \ref{figpk} 
is still apparent in the combined analysis.
The diminished significance in the combined analysis
is to be expected since this
bump is entirely due to the Lyman-$\alpha$ data 
and is not detectable in the CMB.

Regarding other parameters in the analysis, 
it is not surprising that we deduce
values very near to the {\it WMAP}  \cite{wmap}
optimum results,
e.g.~ $n_s = 0.96 \pm 0.01$, $\sigma_8 = 0.88 \pm 0.08$.

\section{Conclusion}
We have analyzed the mass and CMB power spectra in the context of a
model for particle creation during inflation.  We find marginal
evidence for excess power in both the CMB and mass fluctuation spectra
consistent with this hypothesis.  The combined CMB + matter power spectrum
would imply an optimum feature at $k_* = 0.17 \pm 0.04~h$\,Mpc$^{-1}$ 
and $A \approx  0.35 \pm 0.10$.  However, given that the CMB results
may be attributed to the SZ effect and that the matter power spectrum
is very uncertain at this scale, we propose that the 
 most likely possibility could be the
feature in the matter power spectrum deduced from the Lyman-$\alpha$
absorption structures at a scale of $k \sim 0.4~h$ Mpc$^{-1}$ and $A
\approx 0.2$.  Either of these features would correspond to the resonant creation
of a particle with $m \approx 1-2 $ M$_{pl}$ and a Yukawa coupling
constant between the fermion species and the inflaton of $\lambda
\approx 0.5-1.0$ for a single fermion species.
 
Obviously there is a need for more precise determinations of the
primordial matter power spectrum on the scale of 0.1 to 10 Mpc.
Indeed, the
Lyman-$\alpha$ forest constraints and method are currently under revision
and new results are expected shortly which may alter the conclusions presented here.
There is also a need for more
precise determinations of the SZ contribution to the CMB on
the scale of $l \approx 2000 - 10000$.  
We further note that the present analysis has neglected the possibility of
nonadiabatic effects.
Based upon previous studies \cite{Sugiyama,Trotta} of nonadiabatic
isocurvature fluctuations on the matter power spectrum and CMB,
we would expect   
that the introduction of non-gaussian isocurvature fluctuations
could add excess power on small angular scales and hence could decrease the
significance of the features identified here.  They would not, 
however, naturally produce the kind of bump of interest here, and
hence would not easily explain such features away.

In spite of these caveats, we conclude that if the present
analysis is correct, this may be one of the first  hints at observational
evidence of new particle physics at the Planck scale.  Indeed, one
expects a plethora of particles at the Planck scale, particularly in
the context of string theory, the leading candidate for a consistent
theory of quantum gravity.  Perhaps, the presently observed power
spectra contain the first suggestion that such particles may have not
only existed in the early universe, but coupled to the inflaton field,
and thereby, left a relic signature of their existence in the
primordial power spectrum.

\acknowledgements
Work at the University of Notre Dame is supported
by the U.S. Department of Energy under 
Nuclear Theory Grant DE-FG02-95-ER40934.
This work has been supported in part by the Grants-in-Aid for
Scientific Research (12047233, 13640313, 14540271) and for
Specially Promoted Research (13002001) 
of the Ministry of Education, Science, Sports, and Culture of Japan.

\end{document}